\documentclass[aps,twocolumn]{revtex4-1}
\usepackage{graphicx}

\begin{document}

\pacs{87.15.A-, 36.20.Ey, 87.15.H-}
\title{Flexible polymer confined inside a cone-shaped nano-channel }

\author{Narges Nikoofard}
\email{nikoofard@kashanu.ac.ir}
\affiliation{Institute of Nanoscience and Nanotechnology, University of Kashan, Kashan 51167-87317, Iran}

\author{Hossein Fazli}
\affiliation{Department of Physics, Institute for Advanced Studies in Basic Sciences (IASBS), Zanjan 45137-66731, Iran}
\affiliation{Department of Biological Sciences, Institute for Advanced Studies in Basic Sciences (IASBS), Zanjan 45137-66731, Iran}
\date{\today}

\begin{abstract}

Nano-scale confinement of polymer in cone-shaped geometries occurs in many experimental situations. A flexible polymer confined in a cone-shaped nano-channel is studied theoretically and using molecular dynamics simulations. Distribution of the monomers inside the channel, configuration of the confined polymer, entropic force acting on the polymer, and their dependence on the channel and the polymer parameters are investigated. The theory and the simulation results are in very good agreement. The entropic force on the polymer that results from the asymmetric shape of the channel is measured in the simulations and its magnitude is found to be significant relative to thermal energy. The obtained dependence of the force on the channel parameters may be useful in the design of cone-shaped nano-channels.

\end{abstract}

\maketitle

\section{Introduction}

Polymer confinement in nano-scale geometries has become a problem of interest in recent years. This phenomenon has applications in the design of nanotechnology devices, such as polymer separation, DNA sequencing, and protein sensing \cite{separation,rant,sequencing,barcode}. It is also a ubiquitous phenomenon in biological environments. Packaging of the viral genome in the capsid \cite{packaging} and its ejection into the host cell \cite{ejection1,ejection2}, translocation of  RNA through nuclear pores and protein translocation across the endoplasmic reticulum  are known examples \cite{alberts}. Advances in the fabrication of nano-structures for polymer confinement have also lead to considerable achievements in polymer physics. Entropic effects on polymers arisen from confinement were first observed in 1999 \cite{entropic-trapping,craighead1,craighead2}.

Polymer translocation is the passage of a polymer through a nano-pore in a membrane. To date, biological nano-channels have been used for DNA sequencing purposes; while solid-state nano-pores are of interest as a tool for single-molecule studies. The two protein channels used for DNA sequencing are $\alpha$-hemolysin and MspA \cite{dekker}. When DNA passes through the cylindrical part of the $\alpha$-hemolysin channel, the ionic current through the channel depends on 10-15 nucleotides that reside inside the channel. However, the protein channel MspA has a conical shape and the ionic current depends on few nucleotides near the cone tip. Indeed, the electric field is focused in the tip of a cone shaped channel, which also pronounces the difference between the ionic current signals received from the four nucleotides \cite{MspA1,MspA2}.

For interaction between polymers and cone-shaped structures, other biological and technological examples can be counted: The HIV-1 capsid which confines its viral genome has a cone-shaped structure \cite{HIV}, coating of the conical nano-pores with DNA is used for tuning their ionic properties \cite{cone-polymer}, and asymmetric conical pores are shown to act as Brownian ratchets for colloids \cite{colloid} and polymers \cite{unpublished}. In addition, there exist cases that the polymer is confined inside a cone as a result of interaction with the surrounding polymers \cite{werner} or a hydrodynamic flow \cite{sakaue2007}.

Polymers confined near surfaces and inside different geometries such as cylindrical and cone-shaped nano-channels, nano-slits and nano-spheres have been studied (Ref. \cite{hoseinpoor} and references therein). Generally, confinement reduces the number of possible configurations of the polymer and causes its entropy to decrease. For a polymer near a surface, confinement leads to entropic forces on the polymer which are measurable with the enhanced sensitivity of current experimental devices (e.g. AFM and optical tweezers) \cite{kardar1,kantor}. Entropic forces are also the main driving force for polymer escape from confined geometries \cite{luijten}. For a polymer in an asymmetric confinement such as a cone, difference in the entropy of the polymer in the two sides results in an entropic force on the polymer toward the larger side \cite{khalilian}.

\begin{figure*}
\includegraphics[scale=0.23]{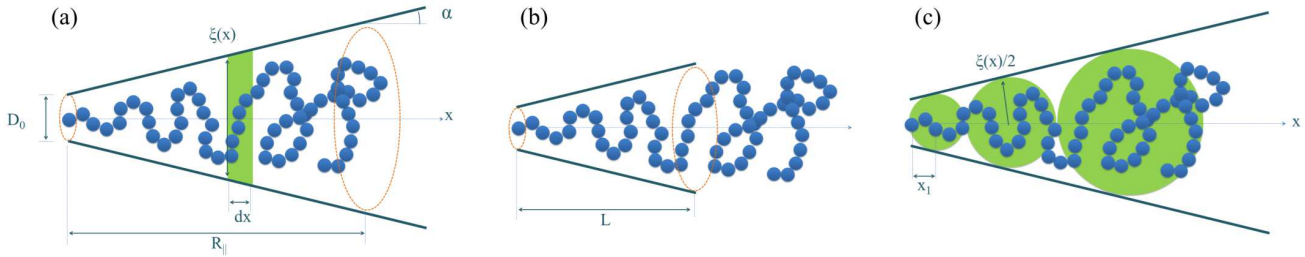}
\caption{(a) Schematics of the polymer inside a long cone-shaped channel (Case I). $\alpha$ is the apex angle and $D_0$ is the opening diameter of the channel, where the polymer end is fixed. The blob size $\xi(x)$ is equal to the local diameter of the channel at position $x$. The polymer is stretched along the channel to size $R_{||}$. The asymmetric shape of the channel exerts an entropic force on the polymer toward the base of the channel.
(b) For a finite channel, a long enough polymer extends to outside the channel (Case II). $L$ defines the channel length along it symmetry axis. The length, the apex angle and the opening diameter of the channel determines the number of monomers that lie inside the channel and the entropic force. (c) A second method to define the blobs. The blobs are tangent to the cone surface and cannot penetrate each other. The first blob is tangent to the beginning of the channel. The number of monomers inside the first blob and the size of the first blob are used to find the limits of applicability of the theory in the cases I and II, respectively.} \label{fig1}
\end{figure*}

In a previous work, the authors studied the translocation of a flexible polymer through a cone-shaped nano-channel, using MD simulations. It was shown that the entropic force from the cone results in a forced polymer translocation. The passage time was shown to have a non-monotonic dependence on the cone angle and a slight dependence on the cone length. A theoretical description was developed for a flexible polymer inside a cone. Small and large cone angles were studies separately, because of the crowding effect in front of the nano-channel in small angles. The number of monomers inside the nano-channel and the entropic force was obtained for a polymer inside a closed cone and inside an open cone-shaped channel. It was shown that the theory can explain the translocation time dependence on the cone angle and length, qualitatively  \cite{khalilian}.

In this paper, we use Molecular Dynamics (MD) simulations to further investigate a polymer confined inside a cone-shaped nano-channel. A flexible polymer that one of its ends is fixed inside the channel is studied by two different sets of simulations. In the first set, the channel is so long that the polymer is completely confined inside it (Case I). Extension of the polymer along the channel axis and the entropic force on the polymer are studied in these simulations. It is seen that the polymer extension along the channel rapidly decreases with the cone angle. However, the force has no considerable dependence on the channel angle and the polymer length. In the second set of simulations, the polymer is so long that more than half of the monomers lie outside the channel (Case II). The number of monomers that remain inside the channel and the force on the polymer are investigated, in this case. The number of monomers inside the channel is found to increase with the channel angle, especially for longer channels. However, the force dependence on the channel length is ignorable. Despite the case I, dependence of the force on the channel angle is considerable, in these simulations. These results on the dependence of the force on the properties of the  cone-shaped channel can be important in the design of nano-pores for practical purposes. The force arisen from the asymmetric shape of the channel on the polymer is around four times the force from thermal fluctuations. So, its effects on the polymer behavior in experimental conditions can be significant which is explained in details in this manuscript. 

The results of MD simulations are also used to check the correctness of the theoretical framework developed in Ref. \cite{khalilian}, quantitatively. It is shown that the four sets of data obtained from the cases I and II can be fitted with the theoretical functions, with only three fit parameters. The theory also explains the distribution of monomers inside the channel. The excellent agreement between the theory and the simulations confirms the strength of the theory in description of a polymer in conical confinement.

The manuscript is organized as follows. In the next section, a brief review of the theory of Ref. \cite{khalilian} and some new notes are presented. In section \ref{method}, the simulation method is described. The simulation results and their agreement with the theory are explained in section \ref{results}. The last section contains a summary of the manuscript, as well as discussions on the correspondence of the results with related experiments and the previously published results.

\section{Theory}    \label{theory}
As case I, consider a flexible polymer that one of its ends is fixed at the tip of an infinitely long cone-shaped channel (Fig. \ref{fig1}(a)). In polymer physics, the blob approach is commonly used to study polymer confinement. In length scales smaller than the blob size, the polymer does not feel that it is confined; while, at length scales larger than the blob size, the confinement is dominant. For a polymer inside a cone, the blob size depends on the position along the axis. The blob size is proportional to the local diameter of the channel; $\xi(x) \sim D_0+2(x+a)\tan\alpha$. $D_0$ and $\alpha$ are the diameter of the beginning of the channel and the apex angle of the channel, respectively. $a$ is the distance along the channel axis between the fixed end of the polymer and the beginning of the channel. Here, $a$ is a mathematical tool and is set equal to zero, after calculations. For clarity, this parameter is not shown in Fig. \ref{fig1}. 

Number of monomers inside each blob is $g(x) \sim \left(\frac{\xi(x)}{b}\right)^{\frac{1}{\nu}}$ . $\nu$ is the Flory exponent and $b$ is the monomer size. The number of monomers inside a volume of thickness $dx$ along the channel is $dn(x) \sim \frac{g(x)}{\xi(x)} dx$. So the linear density of the monomers along the channel axis $\lambda(x) = \frac{dn(x)}{dx}$ becomes
\begin{equation} \label{density}
\lambda(x) \sim \frac{1}{b}\left(\frac{\xi(x)}{b}\right)^{\frac{1}{\nu}-1}.
\end{equation}
Accordingly, the linear density of the monomers  changes with the local diameter of the channel to the power 0.7. 

The polymer extension along the channel, $R_{||}$, is obtained by equating the integral over the number density of the monomers with the total number of monomers of the polymer; $N \sim \int_0^{R_{||}} \lambda(x) dx$. The relation between $N$ and $R_{||}$ becomes
\begin{equation} \label{N}
N \sim \frac{b^{-\frac{1}{\nu}}}{\tan\alpha} \left[\left(D_0+2(a+R_{||})\tan\alpha\right)^{\frac{1}{\nu}}-\left(D_0+2a\tan\alpha\right)^{\frac{1}{\nu}}\right].
\end{equation}

On the other hand, the free energy of confining the polymer is calculated from $\frac{F}{k_BT} \sim \int_0^{R_{||}} \frac{dx}{\xi(x)}$ 
\begin{equation} \label{FE}
\frac{F}{k_BT} \sim \frac{1}{\tan\alpha} \left[\ln\left(D_0+2(a+R_{||})\tan\alpha\right)-\ln\left(D_0+2a\tan\alpha\right)\right].
\end{equation}

The entropic force on the polymer originates from the tendency of the polymer toward the channel base, where it has more space available and larger entropy. As a result, the force is proportional to the changes in the free energy of the polymer when the polymer moves toward the channel base. For a given value of $N$, the force is calculated from the derivative of the free energy with respect to the position of the fixed end of the polymer, $a$. Equation \ref{FE} describes the free energy as a function of the polymer extension $R_{||}$, which is itself a function of the parameter $a$. Considering these points, the force becomes $\frac{f}{k_BT} \sim -\left(1+\left(\frac{\partial R_{||}}{\partial a}\right)_{a=0}\right)\frac{1}{D_0+2R_{||}\tan\alpha}+\frac{1}{D_0}$. The derivative of $R_{||}$ with respect to $a$ is found from eq. \ref{N}. After calculation of the derivatives, $a$ is set equal to zero. The force is obtained as \cite{khalilian}
\begin{equation} \label{fL}
\frac{f}{k_BT} \sim \frac{1}{D_0}\left[1-\left(\frac{D_0}{D_0+2R_{||}\tan\alpha}\right)^{\frac{1}{\nu}}\right]
\end{equation}
 
Constant $B$ is multiplied into equation \ref{N} to convert it into equality.
Then, this equation is rearranged to give the polymer extension $R_{||}$ as a function of the total length of the polymer $N$ and the channel angle $\alpha$ and tip diameter $D_0$;
\begin{equation} \label{L}
\tilde{R}_{||}\tan\alpha =\left[\left(\frac{N}{B}\tan\alpha+\tilde{D}_0^{\frac{1}{\nu}}\right)^{\nu}-\tilde{D}_0\right].
\end{equation}

$A$ is introduced as constant of proportionality into equation \ref{fL}.
Then, equation \ref{L} is substituted into equation \ref{fL} to obtain the force as a function of the polymer and the channel parameters;
\begin{equation} \label{fN}
\tilde{f} = \frac{A}{\tilde{D}_0} \frac{\frac{N}{B}\tan\alpha}{\frac{N}{B}\tan\alpha+\tilde{D}_0^{\frac{1}{\nu}}}.
\end{equation}
The tilde sign shows that the lengths and the force are scaled with $b$ and $\frac{k_BT}{b}$, respectively.

For case II, consider the polymer confined inside a cone-shaped channel of finite length $L$ (the length of the cone-shaped channel along its symmetry axis, see Fig. \ref{fig1}(b)). Trivially, some of the monomers remain outside the channel, for long enough polymers. So, the channel length determines the force and the number of monomers of the polymer that are inside the channel. In this case, $N$ represents the number of monomers inside the channel. Indeed, the monomers outside the channel do not contribute to the entropic force on the polymer, because they do not feel any confinement. In other words, free-energy of the polymer segment outside of the channel does not depend on $a$. 
Taking $a=0$ and substituting $L$ for $R_{||}$, equation \ref{N} describes the number of monomers inside the channel versus the channel parameters $L$, $\alpha$ and $D_0$;
\begin{equation} \label{N2}
N\tan\alpha = B \left[\left(\tilde{D}_0+2\tilde{L}\tan\alpha\right)^{\frac{1}{\nu}}-\tilde{D}_0^{\frac{1}{\nu}}\right].
\end{equation}
The force exerted on the polymer as a function of the channel parameters is obtained from equation \ref{fL} ($R_{||}$ is replaced by $L$);
\begin{equation} \label{fL2}
\tilde{f} = \frac{A}{\tilde{D}_0}\left[1-\left(\frac{\tilde{D}_0}{\tilde{D}_0+2\tilde{L}\tan\alpha}\right)^{\frac{1}{\nu}}\right].
\end{equation}
$A$ and $B$ are the previously introduced proportionality constants.

There is another method to define the confinement blobs for a polymer inside a cone-shaped channel \cite{khalilian} (Fig. \ref{fig1}(c)). The blobs can be defined as spheres that are tangent to the internal surface of the channel and cannot penetrate each other. The size of the blob at each point becomes $\xi(x) = D_0\cos\alpha + 2x\sin\alpha$. The first blob is restricted to be tangent to the beginning of the channel, $x_1 = \frac{\xi(x_1)}{2}$. Using these two relations, the size of the first blob is obtained as $\xi(x_1) \sim \frac{D_0\cos\alpha}{1-\sin\alpha} $. The number of monomers inside the first blob is obtained $g(x_1) \sim \left(\frac{\tilde{D_0}\cos\alpha}{1-\sin\alpha} \right)^{\frac{1}{\nu}}$. $\xi(x_1)$ and $g(x_1)$ can be used to find the limits of applicability of the  theory of a polymer confined inside a cone-shaped nano-channel in the two cases.
If the total number of the monomers is smaller than $g(x_1)$ in case I, the infinite channel would have no confining effect on the polymer. In case II, $\xi(x_1)$ gives the shortest length of the cone-shaped channel that exerts an entropic force on the polymer.
For example, with $D_0=1.4b$ and $\alpha=50\,^{\circ}$, one obtains $g(x_1)=10$ and $\xi(x_1)=4b$.

\section{Simulation method} \label{method}

Coarse-grained MD simulations using ESPResSo \cite{espresso} are employed to study a flexible polymer confined inside a cone-shaped nano-channel. The polymer is modeled by a bead-spring chain. The monomers interact with each other and the channel walls, via the shifted and truncated Lennard-Jones potential $U_{LJ}=4\epsilon\left[\left(\frac{b}{r}\right)^{12}-\left(\frac{b}{r}\right)^{6}+\frac{1}{4}\right]$. The cut-off radius for the potential is $2^{\frac{1}{6}}b$. $\epsilon$ and $b$ are the energy and length scales of the interaction, and $r$ is the distance between the monomers.

Adjacent monomers along the polymer are attached by the FENE potential $U_{FENE}=-\frac{1}{2}KR_0^2\ln\left[1-\left(\frac{r}{R_0}\right)^2\right]$. $K=100\frac{\epsilon}{b^2}$ and $R_0=1.5b$ are the spring constant and the maximum distance between the adjacent monomers.  The simulations are performed under constant temperature $T=1.0\frac{\epsilon}{k_B}$ using the langevin thermostat with the damping constant $1.0\tau_{MD}^{-1}$ \cite{thermostat}. $\tau_{MD}=b\sqrt{\frac{\epsilon}{m}}$ is the time unit of the simulations, where $m$ is the mass of monomers. Equations of motion are integrated using the Velocity-Verlet algorithm, with a time step equal to $0.01\tau_{MD}$.

At the beginning of the simulations, the polymer is arranged on the cone axis. The first monomer at the tip of the channel is fixed during the simulations. All simulations are continued for $5N^2$ time units. The polymer relaxation time is of the order of the polymer length to the power two. As a result, averages of different quantities are measured after the time $N^2$. These quantities are calculated at each time unit (or equally after each 100 time steps) of the simulations. Totally, $4N^2$ configurations of the system are used to calculate the averages.
Prior to calculation of the total average of the force, moving-average is used to reduce the noise in the force data. The noise results from random motions of the polymer on short time scales that the polymer does not feel the confinement. To this end, the force value  at each time is replaced by the average of the force over a time interval starting from the specified time and spanning over $\frac{1}{5}$ of the relaxation time of the polymer. 
The final error-bars are smaller than the size of the symbols, in all plots. 

The smallest diameter of the channel $D_0$ is taken equal to $1.4b$. The channel angle is changed from $1\,^{\circ}$ to $50\,^{\circ}$.
Two sets of simulations are performed. In case I, the channel length is taken to be two times the total length of the polymer. Three different values for the number of monomers of the polymer, 100, 200 and 300 are examined. In case II, the number of monomers of the polymer is determined such that more than half of the monomers lie outside the channel. Test simulations show that further increasing the total number of monomers do not affect the results in this case.
Channel lengths $10b$, $15b$ and $20b$ are tested. In the first case, the radius of gyration and the end-to-end vector of the polymer parallel to the channel axis are calculated. In the second case, the number of monomers inside the channel is counted. 
To find the entropic force on the polymer, sum of the Lennard-Jones forces that the monomers exert on the channel wall is calculated. The result of the force summation is then averaged over time. It is observed that only the component of the average force parallel to the channel axis is nonzero. 
The average entropic force exerted by the channel to the polymer is equal in magnitude but opposite in direction to the above-mentioned forces.

\begin{figure}
\includegraphics[scale=0.8]{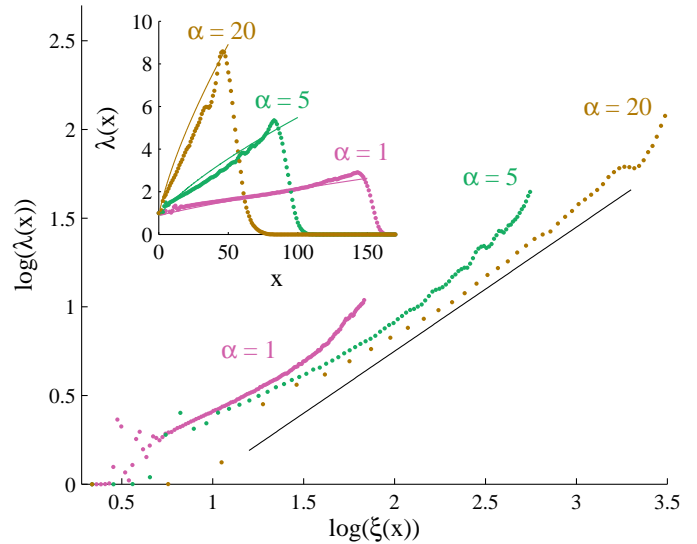}
\caption{Log$_{\text{10}}$-log$_{\text{10}}$ plot of the simulation results for linear density of the monomers along the channel axis versus local diameter of the channel. The simulation results are shown with dots. The polymer has 300 monomers and three different values for the apex angle $1\,^{\circ}$, $5\,^{\circ}$ and $20\,^{\circ}$ are tested.  Data points corresponding to some of the monomers of the two polymer ends are not shown, to remove the ends effect. According to the theory, the linear density of the monomers along the channel axis is proportional to the local diameter of the channel to the power 0.7 (Eq. \ref{density}). The slope of the solid line is 0.7 for comparison with the theory. 
Inset: Simulation results for the linear density of the monomers along the channel axis. Solid lines show the theoretical prediction; $\lambda(\tilde{x})=P(\tilde{D}_0+2\tilde{x}\tan(\alpha))^{\frac{1}{\nu}-1}$. $\tilde{D}_0=1.4$ and $\nu=0.588$ are used and the proportionality constant $P$ is taken as the fit parameter. For all the curves, $P=0.7$ is obtained.} \label{distlc}
\end{figure}

\begin{figure}
\includegraphics[scale=0.8]{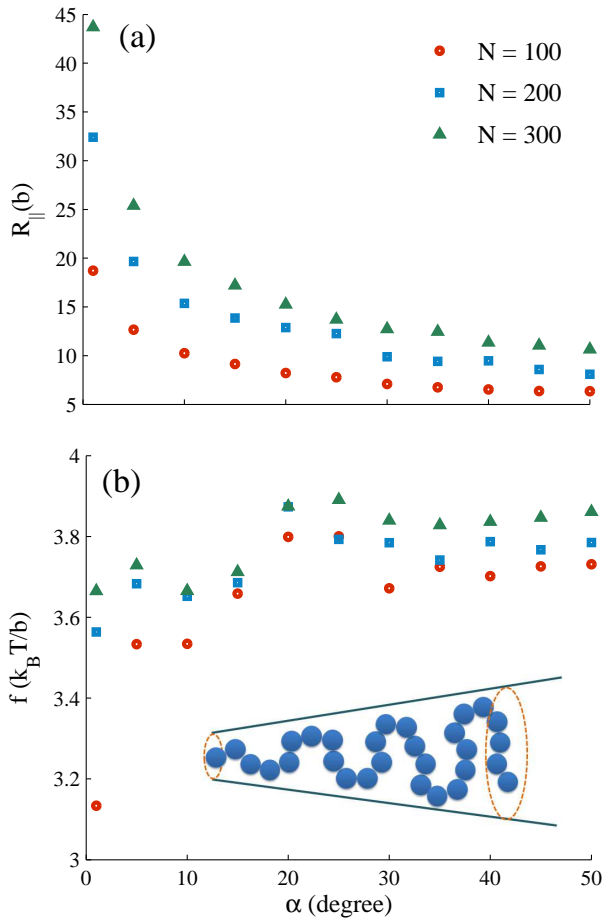}
\caption{Simulation results for a polymer inside a long channel (Case I).   Three different polymer lengths N = 100, 200 and 300 are examined. In simulation of each polymer, the length of the channel is taken equal to two times of the contour length of the polymer. The radius of gyration of the polymer along the channel axis (a) and the entropic force on the polymer (b) are shown versus the channel angle. Although the radius of gyration of the polymer changes rapidly with the channel angle, the entropic force has a weak dependence on the channel angle. The force also depends weakly on the polymer length. This shows that the entropic force originates mainly from the narrow tip of the channel.  A schematic of the polymer inside the long channel is also shown in the panel (b). } \label{lcone}
\end{figure}

\section{Simulation results} \label{results}

\subsection*{Case I: Long channel}
The linear density of the monomers along the channel axis obtained from simulations of a polymer containing 300 monomers is shown in the inset of Fig. \ref{distlc}. Three different angles for the channel $1\,^{\circ}$, $5\,^{\circ}$ and $20\,^{\circ}$ are examined. 
Log-log plot of the linear density of the monomers versus the local diameter of the channel is shown in the main panel of Fig. \ref{distlc}. Data points related to the two ends of the chain are eliminated. The slope of the solid line is 0.7, the exponent predicted by Eq. \ref{density} of the theory. The data  is irregular at small values of the local diameter affected by the fixed end of the polymer. 

\begin{figure}
\includegraphics[scale=0.8]{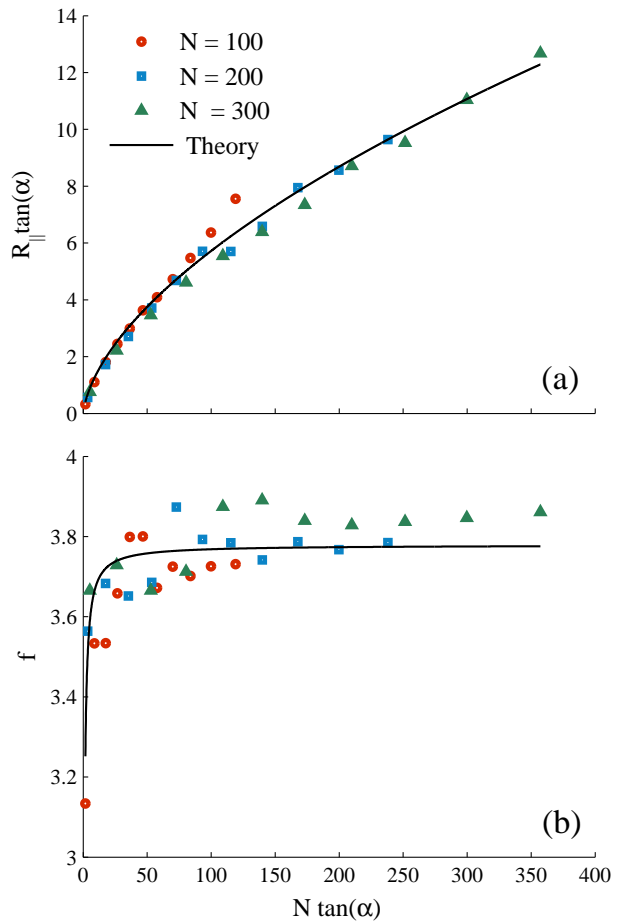}
\caption{Simulation results for a polymer inside a long channel (Case I). Data points of the previous figure are shown in rescaled axes, according to equations \ref{L} and \ref{fN} of the theory. 
As can be seen, data points related to different polymer lengths follow a master curve in the rescaled axes. This shows a good agreement between the theory and the simulation results.
The solid lines show equations \ref{L} and \ref{fN} in panels (a) and (b), respectively. The values $\nu=0.588$, $D_0=1.4b$, $A=5.29$ and $B=0.35$ are used in the equations. The values of the constants $A$ and $B$ are obtained from fitting the theory to the simulation results of case II. Another constant $C=3.5$ is multiplied in the radius of gyration in panel (a). This constant is equal to the ratio of the end-to-end vector of the polymer to its radius of gyration, from the simulation results. } \label{lcone2}
\end{figure}

The radius of gyration of the polymer parallel to the channel axis versus the channel angle is shown in Fig. \ref{lcone}(a). The channel angle is changed from $1\,^{\circ}$ to $50\,^{\circ}$. Three different polymer lengths N = 100, 200 and 300 are examined. The radius of gyration decreases rapidly with the channel angle. Its dependence on the polymer length is considerable in small angles. However, the difference between the three polymers becomes negligible in larger angles.
The entropic force on the polymer versus the channel angle is shown in Fig. \ref{lcone}(b). Despite the radius of gyration of the polymer, the force has a week dependence on the channel angle. Although, the force on the longer polymer is larger for all angles, there is only a slight difference between the forces acting on the polymers of different lengths. This is because the entropic force on the polymer is exerted mainly by the narrow parts of the channel. Overall, the force is around $3.8k_BT/b$ for different channel angles and polymer lengths. Indeed, the magnitude of the entropic force mainly depends on the diameter of the channel in its tip side, $D_0$. The important note is that this entropic force is larger than $k_BT$ and its effect can be considerable in practical situations.

Equations \ref{L} and \ref{fN} describe the polymer in the long channel corresponding to case I. These equations give the polymer extension along the channel and the force on the polymer as a function of the channel angle and the polymer length. It can be deduced from these equations that if $R_{||}\tan\alpha$ and $f$ are plotted versus $N\tan\alpha$, all data points would fall on a master curve. Figures \ref{lcone2}(a) and \ref{lcone2}(b) show the previous figures of case I, with modified axes. It is observed that the curves related to different polymer lengths collapse onto a single curve, according to one's expectation.

\subsection*{Case II: Long polymer}
\begin{figure}
\includegraphics[scale=0.8]{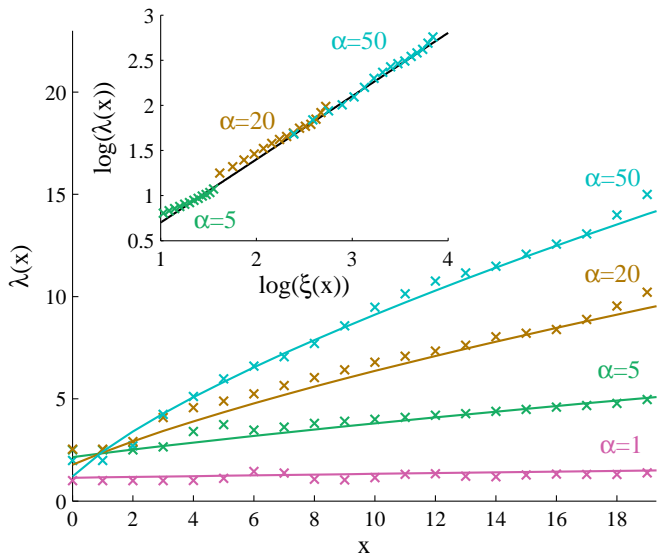}
\caption{Simulation results for linear density of the monomers along the channel axis. The channel length is $L=20b$ and four different channel angles $1\,^{\circ}$, $5\,^{\circ}$, $20\,^{\circ}$ and $50\,^{\circ}$ are examined. Solid lines show the theoretical prediction; $\lambda(\tilde{x})=P(\tilde{D}_0+2\tilde{x}\tan(\alpha))^{\frac{1}{\nu}-1}$. $\tilde{D}_0=1.4$ and $\nu=0.588$ are used and the proportionality constant $P$ is taken as the fit parameter. The obtained values for $P$ change from 1.7 to 0.9, with increasing the apex angle.
Inset: Log$_{\text{10}}$-log$_{\text{10}}$ plot of the simulation results for the scaled linear density of the monomers versus the local channel diameter away from the tip of the channel. The value of the linear density at each angle is divided by the parameter $P$. As is seen, all data collapse on a single line. The black solid line has the slope of 0.7 in agreement with Eq. \ref{density} of the theory. Data points related to $\alpha=1\,^{\circ}$ are not shown, considering the nearly constant value of the local diameter along the channel axis. } \label{distlp}
\end{figure}

Simulation results for linear density of the monomers along the channel axis are sketched in Fig. \ref{distlp}, for monomers inside the channel. Four different angles for the channel $1\,^{\circ}$, $5\,^{\circ}$, $20\,^{\circ}$ and $50\,^{\circ}$ are examined. The log$_{\text{10}}$-log$_{\text{10}}$ plot of the scaled linear density versus the local diameter for points away from the tip of the channel is shown in the inset of Fig. \ref{distlp}. The slope is close to the theoretical value, 0.7.

\begin{figure}
\includegraphics[scale=0.8]{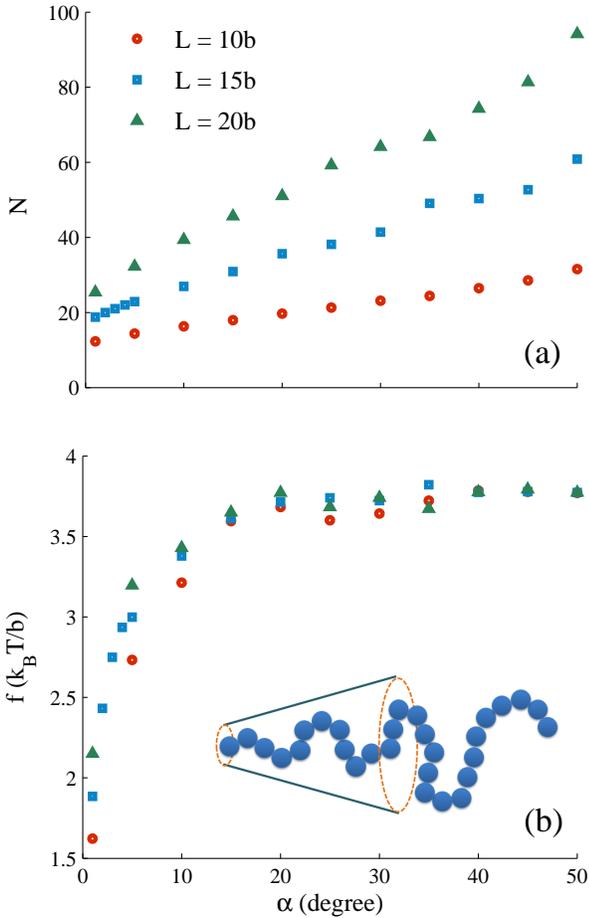}
\caption{Simulation results for a long polymer in a cone-shaped channel (Case II).  Three different channel lengths $L=10b$, $15b$ and $20b$ are tested. Total number of monomers is taken such that more than half of the polymer remains outside the channel.
The number of monomers of the polymer that are inside the channel (a) and the entropic force on the polymer (b) are shown versus the channel angle. The number of monomers inside the channel depends on both the channel length and angle. However, the entropic force  increases and plateaus with the channel angle and does not depend on the channel length. A schematic of the long polymer and the channel is also shown in panel (b).} \label{lpolymer}
\end{figure}

The number of monomers that are inside the channel is plotted in Fig. \ref{lpolymer}(a). The number of monomers inside the channel increases with the channel angle. This increase is especially considerable for the longer channel with $L=20b$.
The entropic force on the polymer versus the channel angle is shown in Fig. \ref{lpolymer}(b). The force dependence on the channel angle is stronger than case I, which is due to the finite length of the channel.  However, despite the growing number of monomers inside the channel (Fig. \ref{lpolymer}(a)), the force on the polymer does not change with the channel length. This again results from the fact that the force originates mainly from the narrow sections of the channel.

In Fig. \ref{lpolymer}(b), the angles between $1\,^{\circ}$ and $5\,^{\circ}$ are also included to compare the results with those of Ref. \cite{khalilian}. It is seen that the force is a monotonic function of the channel angle. This contradicts with functionality of the translocation time of a polymer through a cone-shaped channel, which is a non-monotonic function of the channel angle. It confirms the assumption of Ref. \cite{khalilian} that this non-monotonic behavior is a consequence of the non-equilibrium nature of the translocation process. As the polymer passes through the channel, the monomers crowd at the channel exit and the cone-shaped channel becomes similar to a closed cone-shaped space, for small apex angles.

\begin{figure}
\includegraphics[scale=0.8]{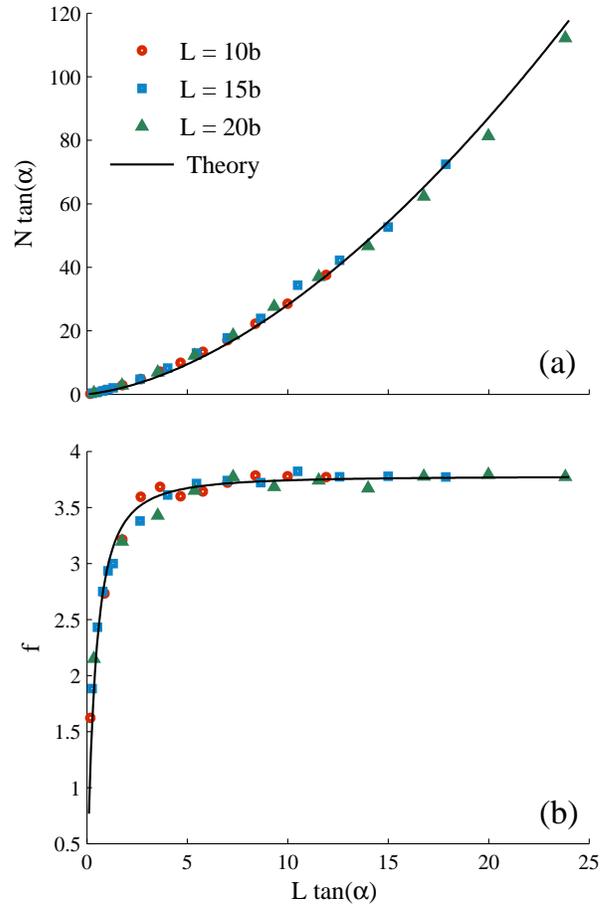}
\caption{Simulation results for a long polymer in a cone-shaped channel (Case II). 
Data points of the previous figure are shown in rescaled axes, according to equations \ref{N2} and \ref{fL2} of the theory. 
As can be seen, data points related to different polymer lengths follow a master curve in the rescaled axes. This shows a good agreement between the theory and the simulation results.
The solid lines are the theoretical fit to the simulation results. Equations \ref{N2} and \ref{fL2} are fitted to the simulation results in panels (a) and (b), respectively. The values $\nu=0.588$ and $D_0=1.4b$ are used, and $A=5.29$ and $B=0.35$ are obtained from the fit. } \label{lpolymer2}
\end{figure}

Equations \ref{N2} and \ref{fL2} describe the long polymer in the channel corresponding to case II. These equations give the number of monomers inside the channel and the force on the polymer versus the channel length and angle. Here, one should plot $N\tan\alpha$ and $f$ versus $L\tan\alpha$ to have all the data points collapsed on a master curve. Figs. \ref{lpolymer}(a) and \ref{lpolymer}(b) are shown with appropriate axes in Figs. \ref{lpolymer2}(a) and \ref{lpolymer2}(b). The data points related to different channel lengths merge into a single curve. This shows an excellent agreement between the theory and the simulations.

\subsection*{Fitting theory to simulation results}

The equations \ref{N2} and \ref{fL2} are used to fit the theory to the simulation results for the case II. The proportionality constants $A$ and $B$ are taken as the fit parameters. The values $\nu=0.588$ and $D_0=1.4$ are assumed, according to the simulation conditions. The theoretical results for the number of monomers inside the channel and the entropic force exerted by the channel are shown in Figs. \ref{lpolymer2}(a) and \ref{lpolymer2}(b). It can be seen that the theoretical function follows the simulation results, very well. The fit parameters are obtained as $A=5.29$ and $B=0.15$.

The values obtained for $A$ and $B$ in case II are substituted in equation \ref{fN}, to describe the force exerted by the cone in case I. The resulting function is sketched in Fig. \ref{lcone2}(b). It is seen that the theory describes the simulation results very well. 
Equation \ref{L} describe the extension of the polymer $R_{||}$ inside the cone-shaped nano-channel. However, the radius of gyration of the polymer $R_g$ is measured in the simulations. These two quantities are proportional to each other, for long channels. As a result, it is necessary to define a third proportionality constant $C$. Equations \ref{L} and $R_{||}=CR_g$ are fitted to the simulation results. The theoretical function is in excellent agreement with the simulation results. The fit parameter is obtained as $C=3.5$. This value is equal to the ratio of the end-to-end vector of the polymer to its radius of gyration, from simulation results. 

\subsection*{Force dependence on the tip diameter}

\begin{figure}
\includegraphics[scale=0.8]{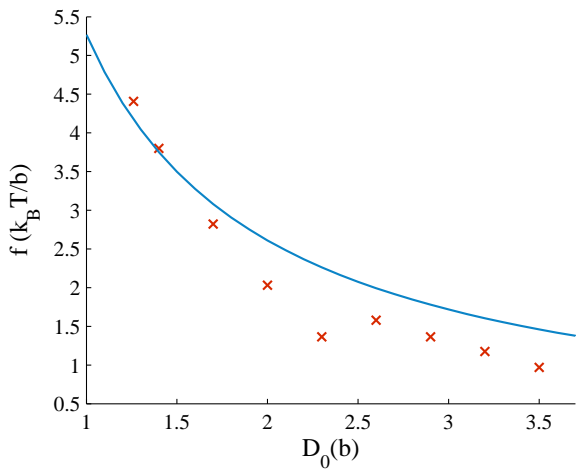}
\caption{Entropic force versus the tip diameter of the channel in case I. The symbols are the simulation results for a polymer with 100 monomers inside a long channel with the angle $\alpha=20\,^{\circ}$. The values $\nu=0.588$, $A=5.29$ and $B=0.35$ are used in Eq. \ref{fN} to obtain the theoretical prediction shown with the solid line.
The force dependence on the tip diameter is considerable, comparing with its dependence on the polymer length and the channel angle shown in Fig. \ref{lcone}(b). A good qualitative agreement can be seen between theory and the simulation data. } \label{force1d}
\end{figure}

In the simulations of case I, the force does not depend on the polymer length. This means that the force originates mainly from the narrow part (the tip side) of the channel. This is also observed in further simulations; by changing $D_0$ from $1.26b$ to $3.5b$, the force drops from $4.4\frac{k_BT}{b}$ to below $\frac{k_BT}{b}$. The simulation results for a polymer with 100 monomers and a channel with $\alpha=20\,^{\circ}$ are shown with symbols in Fig. \ref{force1d}. This strong dependence of the force on the tip diameter of the channel can be independently obtained from eq. \ref{fN} of the theory. The force however depends weakly on the other parameters of the system in case I (the channel angle and the polymer length), according to both the theory and the simulations. For case II, the theory also predicts a sharp change of the force with the tip diameter of the channel.

\section{Summary and discussion} \label{discuss}

In summary, we have studied polymer confinement inside a cone-shaped channel, theoretically and using MD simulations. The problem is considered in two cases: (I) the channel is longer than the polymer, and (II) the polymer is longer than the channel. The polymer extension along the channel and the number of monomers of the polymer that lie inside the channel are examined in cases I and II, respectively. The entropic force originated from the asymmetric shape of the channel is investigated in both cases. It was shown that the polymer extension along the channel and the number of monomers inside the channel change rapidly with the channel angle and length. However, the force from the channel in case I does not depend on the polymer length and the channel angle. In case II, the entropic force depends on the channel angle for small values of the apex angle. In both cases, the force depends strongly on the opening diameter of the channel.

It is also instructive to have a comparison between the results presented here and those of Ref. \cite{khalilian}. Polymer translocation through a cone-shaped nano-channel under no external forces has been consiered in Ref. \cite{khalilian}, while the entropic force and the static properties of the polymer are studied in equilibrium condition in the present study. In Ref. \cite{khalilian}, the entropic force from the nano-channel is the driving force for the translocation process. 
According to the present study, the entropic force increases and saturates with the apex angle. Because the entropic force is the driving force of the translocation process, one expects the translocation time to decrease and then plateau with the angle. This trend is generally observed in the results of ref. \cite{khalilian}. The translocation time reduces sharply from that of a cylindrical channel in small apex angles.
However, a slight local maximum is observed afterwards in the middle values of the apex angle. This is related to the non-equilibrium crowding of  the monomers behind the nano-channel at forced polymer translocation. The crowding effect increases with the driving force and decreases with the area of the channel exit. According to the present study, the entropic force reaches to its highest value at $\alpha=20\,^{\circ}$ and then plateau. The local maximum of the translocation time is observed around this angle and then there is a plateau in larger angles. 
In the previous study, the translocation time had a weak dependence on the channel length, in agreement with the results of the present study. 

According to the present simulation results, the force from the channel has a maximum around $4\frac{k_BT}{b}$, where $b$ is the monomer size. If we consider the translocating polymer to be single-stranded DNA (ssDNA), the size of the coarse-grained monomers is around $1nm$ \cite{rapid}. So, the entropic force from the cone can be as large as $16pN$, which can be considerable in practical situations. 

In experiments, the MspA protein channel has an apex angle around $10\,^{\circ}$ and an opening diameter of $D_0=1.2b$ \cite{trends}. ssDNA is usually entered into the MspA through its base, using an applied electric voltage. 
It is possible to immobilize the strand inside the nano-channel, using a hairpin construct at one end \cite{MspA1} or a DNA polymerase \cite{MspA2}. 
The entropic force on the immobilized ssDNA from the MspA channel can be calculated theoretically from eq. \ref{fL2}, which is approximately equal to $16pN$. 
Although a detailed knowledge of the electric field inside the cone-shaped channel is required to find the electric force, it can be estimated using the relation $F_{elc}=q\frac{\Delta V}{\ell}$.  
Most of the applied electric voltage $\Delta V$ falls on the narrow constriction of the cone-shaped nano-channel and one can assume an effective length for the nano-channel $\ell=2nm$.
$q$ is the total charge of the nucleotides that are inside this effective length. Each nucleotide has 0.3nm length and one electron charge, $e$. 
The applied voltage difference that gives an electric force equal to the entropic force from the channel can be obtained as $\Delta V=30mV$. 

Experimentally, the threshold voltage required to keep ssDNA inside the MspA channel can be measured. The threshold voltage depends on the entropic contributions both from the nano-channel and the membrane containing the nano-channel. An immobilized polymer inside a nano-channel can be divided into three segments: the segment inside the channel and the two arms outside the channel. Each of the two outside arms of the polymer is nearly fixed on the wall containing the nano-channel. The wall reduces the available configurations for the polymer and thus exerts an entropic force on the polymer \cite{kardar2}.
Comparison between the threshold voltage of MspA and that of a cylindrical nano-channel can be used to separate the two entropic contributions. ssDNA fixed inside a cylindrical nano-channel (such as $\alpha$-hemolysin) feels only the entropic force from the wall and can be used as a reference to find the entropic force from the MspA channel.

In the experiments of ssDNA translocation through MspA, the voltage difference  $\Delta V=180mV$ is often used which gives an electric force equal to $96pN$ on the strand. The polymer is often entered from the base of the cone-shaped channel, to achieve higher capture rates \cite{entry,capture}. In this condition, the entropic force from the channel acts against the direction of the polymer entry and can reduce the capture rate. However, non-equilibrium effects are determining in polymer translocation. The polymer becomes stretched when it enters the pore \cite{sakaue2007,farahpour} and the monomers crowd close to the pore after the passage. As ssDNA enters from the base of the MspA channel, crowding does not alter the entropic force but stretching can reduce interaction of the polymer with the channel walls and decrease the entropic force.

Finally, the theory and the simulation results for the polymer configuration and the entropic force on the polymer inside a cone-shaped channel are in good agreement. The simulation results are well fitted with theoretical functions. It was shown that the entropic force from the channel can be significant in ssDNA translocation through MspA channel. In these experiments, the entropic force depends on the angle and the tip diameter of the channel, and not on the channel length. It means that changing the angle of the cone-shaped channel may have a determining role in polymer translocation but changing the channel length has no noticeable effects. This point can be considered in the design of cone-shaped channels through protein engineering \cite{MspA2}, or through solid state methods in fabricating solid-state cone-shaped channels \cite{solid}. In some experiments, the polymers are used to alter transport properties of cone-shaped nano-channels \cite{cone-polymer}. Distribution of monomers and the polymer extension inside the cone-shaped nano-channel are important in these experiments. The obtained results give a detailed view on these quantities.


\begin{thebibliography}{2}

\bibitem{separation}
J. Fu, R. B. Schoch, A. L. Stevens, S. R. Tannenbaum, and J. Han, \emph{Nature Nanotech.}, 2007, \textbf{2}, 121.

\bibitem{rant}
R. Wei, V. Gatterdam, R. Wieneke, R. Tampe, and U. Rant, \emph{Nature Nanotech.}, 2012, \textbf{7}, 257.

\bibitem{sequencing}
D. P. Hoogerheide, B. Lu, and J. A. Golovchenko, \emph{ACS Nano}, 2014, \textbf{8}, 7384.

\bibitem{barcode}
R. Marie \emph{et al.}, \emph{Proc. Natl. Acad. Sci. USA}, 2013, \textbf{110}, 4893.

\bibitem{packaging}
Z. T. Berndsen, N. Kellera, S. Grimes, P. J. Jardine, and D. E. Smith, \emph{Proc. Natl. Acad. Sci. USA}, 2014, \textbf{111}, 8345.

\bibitem{ejection1}
I. J. Molineux and D. Panja, \emph{Nature Reviews Microbiology}, 2013, \textbf{11}, 194.

\bibitem{ejection2}
D. Marenduzzo, C. Micheletti, E. Orlandini, and D. W. Sumners, \emph{Proc. Natl. Acad. Sci. USA}, 2013, \textbf{110}, 20081.

\bibitem{alberts}
B. Alberts, A. Johnson, J. Lewis, M. Raff, K. Roberts, and P. Walter, \emph{Molecular Biology of the Cell}, Garland Science, New York, 2008.

\bibitem{entropic-trapping}
J. Han, S. W. Turner, and H. G. Craighead, \emph{Phys. Rev. Lett.}, 1999, \textbf{83}, 1688.

\bibitem{craighead1}
S.W. P. Turner, M. Cabodi, and H. G. Craighead, \emph{Phys. Rev. Lett.}, 2002, \textbf{88}, 128103.

\bibitem{craighead2}
C. H. Reccius, J. T. Mannion, J. D. Cross, and H. G. Craighead, \emph{Phy. Rev. Lett.}, 2005, \textbf{95}, 268101.

\bibitem{dekker}
G. F. Schneider and C. Dekker, \emph{Nature Biotech.}, 2012, \textbf{30}, 326.

\bibitem{MspA1}
I. M. Derrington, T. Z. Butler, M. D. Collins, E. Manrao, M. Pavlenok, M. Niederweis, and J. H. Gundlach, \emph{Proc. Natl. Acad. Sci. U.S.A.}, 2010, \textbf{107}, 16060.

\bibitem{MspA2}
E. A Manrao, I. M. Derrington, A. H. Laszlo, K. W. Langford, M. K. Hopper, N. Gillgren, M. Pavlenok, M. Niederweis, and J. H. Gundlach, \emph{Nature Biotech.}, 2012, \textbf{30}, 349.

\bibitem{HIV}
G. Zhao \emph{et al.},  \emph{Nature}, 2013, \textbf{497}, 643.

\bibitem{cone-polymer}
G. Nguyen, S. Howorka, Z. S. Siwy, \emph{J. Membrane Biol.}, 2011, \textbf{239}, 105.

\bibitem{colloid}
S. Matthias and F. Muller, \emph{Nature}, 2002, \textbf{424}, 53.

\bibitem{unpublished}
N. Nikoofard and M. Mikani, unpublished work.

\bibitem{werner}
M. Werner and J. U. Sommer, \emph{Eur. Phys. J. E}, 2010, \textbf{31}, 383.

\bibitem{sakaue2007}
T. Sakaue, \emph{Phys. Rev. E}, 2007, \textbf{76}, 021803.

\bibitem{hoseinpoor}
N. Nikoofard, S. M. Hoseinpoor, M. Zahedifar, \emph{Phys. Rev. E}, 2014, \textbf{90}, 062603.

\bibitem{kardar1}
M. F. Maghrebi, Y. Kantor, and M. Kardar, \emph{Eur. Phys. Lett.}, 2011, \textbf{96}, 66002.

\bibitem{kantor}
Y. Hammer and Y. Kantor, \emph{Phys. Rev. E}, 2014, \textbf{89}, 022601.

\bibitem{luijten}
A. Cacciuto and E. Luijten, \emph{Phys. Rev. Lett.}, 2006, \textbf{96}, 238104.

\bibitem{khalilian}
N. Nikoofard, H. Khalilian, and H. Fazli, \emph{J. Chem. Phys.}, 2013, \textbf{139}, 074901.

\bibitem{espresso}
H. J. Limbach, A. Arnold, B. A. Mann, and C. Holm, \emph{Comp. Phys. Communications}, 2006, \textbf{174}, 704.

\bibitem{thermostat}
N. Goga, A. J. Rzepiela, A. H. de Vries, S. J. Marrink, and H. J. C. Berendsen, \emph{J. Chem. Theory Comput.}, 2012, \textbf{8}, 3637.

\bibitem{rapid}
N. Nikoofard and H. Fazli, \emph{Phys. Rev. E}, 2011, \textbf{83}, 050801(R).

\bibitem{trends}
S. W. Kowalczyk, T. R. Blosser and C. Dekker, \emph{Trends in Biotechnology}, 2011, \textbf{29}, 607.

\bibitem{kardar2}
J. Chuang, Y. Kantor, and M. Kardar, \emph{Phys. Rev. E}, 2001, \textbf{65}, 011802.

\bibitem{entry}
N. Nikoofard and H. Fazli, \emph{Phys. Rev. E}, 2012, \textbf{85}, 021804.

\bibitem{capture}
M. Mihovilovic, N. Hagerty, and D. Stein, \emph{Phys. Rev. Lett.}, 2013, \textbf{110}, 028102.

\bibitem{solid}
C. C. Harrell, Z. S. Siwy, and C. R. Martin, \emph{Small}, 2006, \textbf{2}, 194.

\bibitem{farahpour}
F. Farahpour, A. Maleknejad, F. Varnik, M. R. Ejtehadi, \emph{Soft Matter}, 2013, \textbf{9}, 2750.

\end{thebibliography}
\end{document}